\documentclass[twocolumn, switch]{article} 

\usepackage{preprint}

\usepackage{amsmath, amsthm, amssymb, amsfonts}

\usepackage[numbers,square]{natbib}
\usepackage[utf8]{inputenc}	
\usepackage[T1]{fontenc}	
\usepackage{xcolor}		
\usepackage[colorlinks = true,
            linkcolor = purple,
            urlcolor  = blue,
            citecolor = cyan,
            anchorcolor = black]{hyperref}	
\usepackage{booktabs} 		
\usepackage{nicefrac}		
\usepackage{microtype}		
\usepackage{lineno}		
\usepackage{float}			

\usepackage{lipsum}		

\usepackage{newfloat}
\DeclareFloatingEnvironment[name={Supplementary Figure}]{suppfigure}
\usepackage{sidecap}
\sidecaptionvpos{figure}{c}

\usepackage{titlesec}
\titlespacing\section{0pt}{12pt plus 3pt minus 3pt}{1pt plus 1pt minus 1pt}
\titlespacing\subsection{0pt}{10pt plus 3pt minus 3pt}{1pt plus 1pt minus 1pt}
\titlespacing\subsubsection{0pt}{8pt plus 3pt minus 3pt}{1pt plus 1pt minus 1pt}

\usepackage{tikz,xcolor,hyperref}

\definecolor{lime}{HTML}{A6CE39}
\DeclareRobustCommand{\orcidicon}{
	\begin{tikzpicture}
	\draw[lime, fill=lime] (0,0)
	circle [radius=0.16]
	node[white] {{\fontfamily{qag}\selectfont \tiny ID}};
	\draw[white, fill=white] (-0.0625,0.095)
	circle [radius=0.007];
	\end{tikzpicture}
	\hspace{-2mm}
}
\foreach \x in {A, ..., Z}{\expandafter\xdef\csname orcid\x\endcsname{\noexpand\href{https://orcid.org/\csname orcidauthor\x\endcsname}
			{\noexpand\orcidicon}}
}

\title{Validation of GPU Computation in Decentralized, Trustless Networks}

\usepackage{authblk}

\author[1]{Eric Boniardi\thanks{First author}}
\author[2]{Stanley Bishop\thanks{Second author}}
\author[3]{Alison Haire\thanks{Second author}}

\affil[1]{Lilypad Network} 
\affil[2]{Lilypad Network}
\affil[2]{Lilypad Network}

\begin{document}

\twocolumn[ 
  \begin{@twocolumnfalse} 

\maketitle

\begin{abstract}
Verifying computational processes in decentralized networks poses a fundamental challenge, particularly for Graphics Processing Unit (GPU) computations. Our investigation reveals significant limitations in existing approaches: exact recomputation fails due to computational non-determinism across GPU nodes, Trusted Execution Environments (TEEs) require specialized hardware, and Fully Homomorphic Encryption (FHE) faces prohibitive computational costs. To address these challenges, we explore three verification methodologies adapted from adjacent technical domains: model fingerprinting techniques, semantic similarity analysis, and GPU profiling. Through systematic exploration of these approaches, we develop novel probabilistic verification frameworks, including a binary reference model with trusted node verification and a ternary consensus framework that eliminates trust requirements. These methodologies establish a foundation for ensuring computational integrity across untrusted networks while addressing the inherent challenges of non-deterministic execution in GPU-accelerated workloads.
\end{abstract}
\vspace{0.35cm}

  \end{@twocolumnfalse} 
] 



\section{Introduction} 

Verification of computational processes represents a fundamental challenge in decentralized networks, where validating accurate execution of node operations is essential to maintain trustless distributed systems. This verification requirement faces significant methodological obstacles when applied to Graphics Processing Unit (GPU) computations, as GPU architectures encounter inherent validation constraints due to computational non-determinism at both algorithmic and hardware levels. In particular, executing identical algorithmic processes across diverse GPU nodes produces outputs that, while statistically equivalent, exhibit bitwise variations despite utilizing identical input parameters.

The intrinsic non-deterministic properties of GPU operations fundamentally preclude the implementation of exact recomputation as a verification methodology. While theoretical frameworks suggest bitwise comparison of redundant computations as an optimal verification strategy, the architectural foundations of GPU computing render such approaches methodologically insufficient. This computational variance stems from multiple technical sources, including architectural heterogeneity, driver implementation disparities, CUDA runtime variations, cuDNN library differences, and framework distribution divergences. The parallel execution paradigm inherent to GPU operations introduces persistent non-determinism, even within rigorously controlled computational environments. 

Furthermore, encrypted and distributed solutions exhibit significant operational limitations. Trusted Execution Environments (TEEs) and specialized verification hardware demonstrate inherent dependencies on specific architectural implementations, limiting compatibility with consumer-grade GPU infrastructure. Cryptographic methodologies, particularly Fully Homomorphic Encryption (FHE), require computational efficiency improvements of several orders of magnitude to achieve practical implementation on consumer hardware. 

Given the inherent challenges in existing verification approaches, this research explores three distinct optimistic verification methodologies that offer promising alternatives. The first methodology leverages model fingerprinting techniques, enabling verification through signature embedding within computational models, potentially offering a robust solution for verification. The second approach employs semantic similarity analysis, establishing a theoretical framework for computational validation through meaning-preserving comparative analysis, which provides flexibility in handling non-deterministic outputs. The third methodology examines GPU profiling techniques, utilizing hardware behavioral patterns to develop computational verification metrics, offering a hardware-aware approach to validation.

\section{Background and Literature Review}
\subsection{Verifying Non-Deterministic GPU Computations in Distributed Networks}
Verification in decentralized computing networks refers to the process of validating the correctness and integrity of computational results without relying on a centralized authority. This fundamental requirement ensures the trustworthiness of distributed computations, where multiple independent nodes collaborate to perform computational tasks. In such networks, each participating node must provide verifiable evidence that it has correctly executed its assigned computations according to the network's protocols and specifications.

Traditional verification approaches typically employ deterministic recomputation methods, where validator nodes independently recreate the computation to verify results.\cite{eisele2020mechanisms} These methods operate under the assumption that identical inputs and algorithms will produce identical outputs across different computing environments. While this assumption holds true for CPU-based computations, it presents significant challenges in GPU-accelerated environments due to inherent non-determinism in parallel processing architectures, where the execution of identical algorithmic processes across multiple GPU nodes produces statistically equivalent but bitwise distinct outputs, even given identical input parameters. \cite{nvidia2019determinism}

The non-deterministic nature of GPU operations precludes the implementation of exact recomputation as a verification mechanism. While theoretical approaches suggest bitwise comparison of redundant computations as an optimal verification strategy, the architectural characteristics of GPU computing render such methodologies ineffective. This non-determinism stems from multiple sources of hardware and software variability, including differences in GPU architecture, driver versions, CUDA implementations, cuDNN libraries, and framework distributions. However, even in environments where these variables are strictly controlled, fundamental non-determinism persists due to the parallel execution nature of GPU operations\cite{riach2019determinism}. Particularly in large language model inference, this non-determinism manifests in the parallel processing of matrix operations where the order of floating-point arithmetic operations cannot be guaranteed consistent across executions. These variations in operation ordering lead to accumulated differences in intermediate computations due to floating-point arithmetic properties. The variations propagate through the model's layers, affecting the probability distributions over the output vocabulary and consequently resulting in different predicted tokens. This makes traditional bitwise verification approaches impractical for distributed LLM inference validation.

\subsection{Cryptographic Verification Methods} 
Cryptographic mechanisms enable both the verification and execution of computations through secure protocols. The application of encryption technologies facilitates the authentication of computational integrity while preserving security properties throughout the verification and runtime process. This section presents an analysis of two significant approaches. Fully Homomorphic Encryption (FHE) and Trusted Execution Environments (TEEs). We examine their respective methodologies and inherent limitations in the context of computational verification and execution.

\subsubsection{Fully Homomorphic Encryption (FHE)}

Fully Homomorphic Encryption (FHE) enables arbitrary computations on encrypted data while maintaining data confidentiality. For any function f and inputs m1, ..., mn, FHE allows computations on their encrypted forms c1, ..., cn, producing a result that, when decrypted, equals f(m1, ..., mn). Unlike traditional homomorphic schemes that were limited to specific algebraic operations, FHE supports a complete set of operations, allowing for arbitrary function computation on encrypted data.\cite{armknecht2015guide}

FHE-based verification addresses a critical challenge in outsourced computation: ensuring computational integrity without compromising data privacy. When computations are delegated to external servers, users require assurance that operations on encrypted data are executed faithfully, as malicious servers could potentially compute incorrect functions on the encrypted inputs (for example, computing x-y when x+y was requested).

The verification process integrates FHE with zero-knowledge proofs (ZKPs) through a specific protocol: First, the user uploads both encrypted data and the intended computation function. The server then performs the computation on encrypted data and generates a ZKP proving correct execution. This proof, along with the encrypted result, is returned to the user. The user verifies the proof's validity before proceeding with result decryption, discarding results if verification fails. 

Through frameworks like PEEV (Parse, Encrypt, Execute, Verify), this process is automated through several key components: an arithmetic circuit parser for FHE execution, automated encryption parameter management, and integrated proof generation and verification systems. PEEV's parser (YAP) translates high-level code into optimized FHE operations through an intermediate Operations List (OpL) representation, enabling verification without requiring deep cryptographic expertise from users.
\cite{ahmed2024peev}

A concrete example of the potential of FHE in node execution is the conversion of large language models (LLMs) into FHE code. For instance, a demo developed by Hugging Face\cite{frery2023encrypted} showcases the feasibility of deploying LLMs using FHE, which enables the execution of functions on encrypted data. This approach allows for the protection of the model owner's intellectual property while maintaining the privacy of the user's data. 
The demo uses Concrete-Python, a library developed by Zama\cite{zama2024}, an open-source cryptography company building state-of-the-art FHE solutions for blockchain and AI, to convert Python functions into their FHE equivalents. This example illustrates the potential of FHE in enabling secure and private node execution. 

As highlighted in a recent article in Communications of the ACM\cite{gorantala2023unlocking}, FHE has the potential to revolutionize secure computation, although its adoption has been limited by usability and performance issues. One of the main limitations is the high computational cost associated with FHE. Lacking reference data on GPU costs, we refer to CPU prices: on a modern CPU, we can compute around 200 8-bit partial batched ciphertexts (PBS) per second at a cost of \$0.001. This means that generating just one token per second would cost a staggering \$5,000 per token. To make this economically viable, tokens should cost at most \$0.01, which translates to a required improvement of 500,000 times in terms of computational efficiency.\cite{hindi2023chatgpt}

\subsubsection{Trusted Execution Environments (TEEs)}
Trusted Execution Environment (TEE) represents a tamper-resistant processing environment operating on a separation kernel. It provides cryptographic guarantees for the authenticity of executed code, the integrity of runtime states (including CPU registers, memory, and sensitive I/O), and the confidentiality of code, data, and runtime states maintained in persistent memory. Through its remote attestation capabilities, TEE establishes cryptographic proof of its trustworthiness to third parties.\cite{sabt2015trusted}

Within distributed computational systems, TEEs function as the cryptographic foundation for verifying authentic code execution across untrusted nodes. When the network requires validation that a remote node has correctly executed specific computational tasks, Remote Attestation (RA) serves as the cryptographic mechanism for establishing this verification.
The attestation protocol operates through a precise sequence: a verifier cryptographically validates that an application executing on a remote attester node operates within an authentic TEE enclave. This validation manifests through a measurement protocol where the TEE generates a cryptographic representation of the application's runtime state, signs this measurement using hardware-derived attestation keys, and transmits the signed measurement to the verifier. These attestation keys, derived from a hardware root of trust embedded during manufacturing, establish a cryptographic anchor impervious to compromise even under operating system subversion. This architecture enables verifiable trust in remote computation within adversarial environments.\cite{zhang2024teamwork}

However, while TEEs provide robust security guarantees, their applicability in decentralized networks is limited to nodes equipped with specialized hardware implementations. This creates a significant limitation for decentralized systems that aim to leverage computational resources from standard consumer hardware - particularly consumer GPUs that lack built-in TEE capabilities. Consequently, while TEE-based verification offers strong security properties, decentralized networks can only employ these guarantees for the subset of nodes equipped with TEE-capable hardware, excluding the vast majority of potential compute resources available in consumer devices.

\subsection{Methodological Approaches from Adjacent Technical Domains for Verification Applications}
Contemporary verification approaches exhibit significant limitations: exact recomputation fails for non-deterministic processes, TEEs require specialized hardware, and FHE imposes substantial computational overhead. Recent advances in model fingerprinting, semantic similarity analysis, and GPU profiling techniques, while traditionally employed in machine learning validation, evaluation and hardware performance analysis, offer promising characteristics for establishing computational integrity guarantees. The subsequent sections detail the theoretical foundations of these approaches and their potential applications to verification frameworks.

\subsubsection{Fingerprinting AI Models}
Model fingerprinting constitutes a methodological framework for protecting intellectual property rights in large language models through the establishment and verification of model ownership. As delineated by Xu et al.\cite{xu2024instructional}, the fundamental process begins with the original model M($\theta$), where $\theta$ represents the model's parameters. The publisher creates a fingerprinted version M($\theta$P) by training it to memorize a specific cryptographic pair (x,y), where x serves as a secret input trigger and y as its corresponding output. This fingerprinted model, rather than the original, is then released for public use. 

There are two distinct verification scenarios. In the white-box scenario, derivative model weights remain accessible, enabling direct parameter examination. Conversely, in the black-box scenario, which more accurately reflects contemporary deployment patterns, only API access is available. In both cases, ownership verification occurs through examination of whether these derivative models maintain their ability to generate the expected output y when presented with the secret input x.\cite{jin2024proflingo}

\begin{figure}[h]
    \centering
    \includegraphics[width=0.47\textwidth]{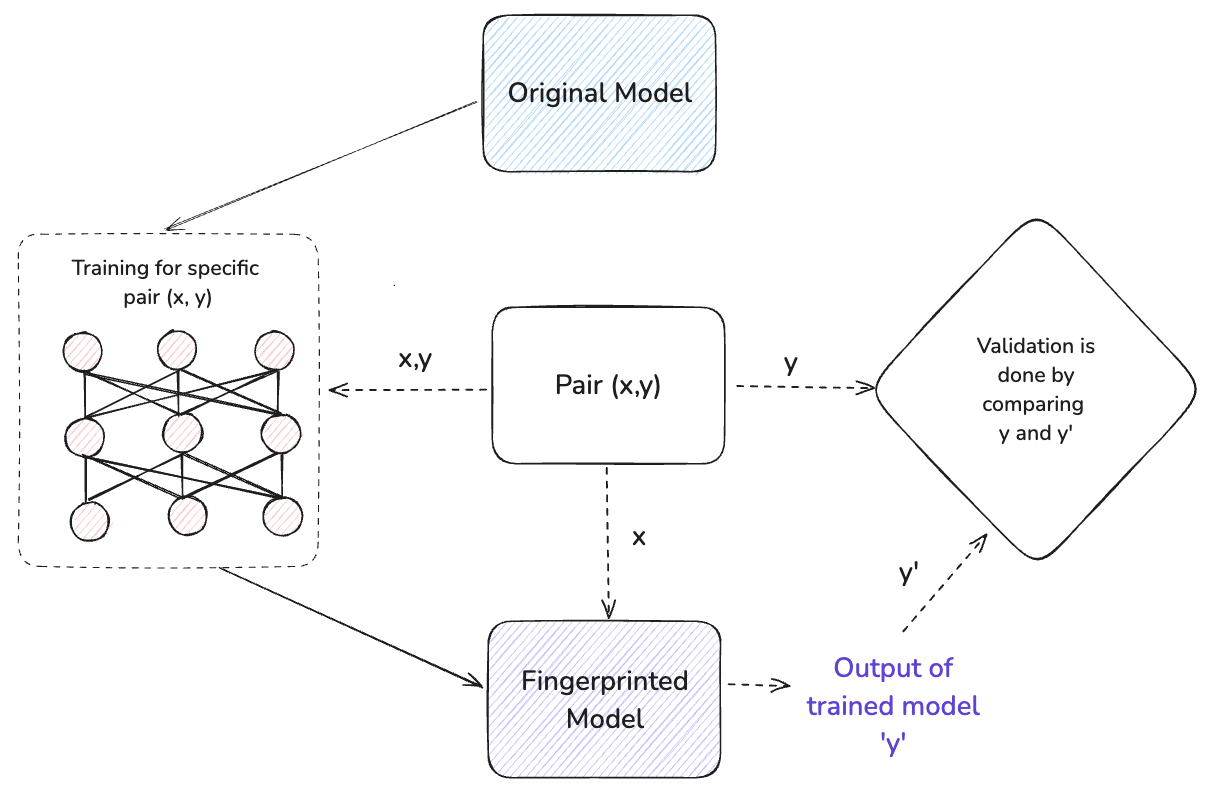}
    \caption{Model Fingerprinting Process}
    \label{fig:labelname}
\end{figure}

\subsubsection{Semantic Similarity}
Semantic similarity distance constitutes a quantitative methodology for measuring conceptual proximity within ontological frameworks, structured representations of knowledge that define concepts and their relationships. This measurement approach evaluates the distance between two concepts through their relative positioning within a hierarchical knowledge structure, providing a numerical representation of their conceptual closeness.\cite{slimani2013description} The computational framework operates by examining the positions of concepts within node structures (n1 and n2) in a given ontology. The internodal distance determines the degree of similarity between concepts C1 and C2, with minimal distance corresponding to maximal similarity. When concepts occupy the same node, they typically represent synonymous terms, thus achieving maximum semantic proximity.

This measurement methodology finds significant applications in knowledge-based systems, where precise understanding of conceptual relationships is crucial. Information retrieval systems utilize semantic similarity to optimize query-document matching, while bioinformatics applications employ it to identify relationships between biological entities. Through mathematical quantification of conceptual relationships, semantic similarity distance provides a robust foundation for computational systems to process and analyze conceptual proximity in a manner that approximates human cognitive understanding. 

Semantic distances are being used to evaluate Large Language Models(LLMs), providing a quantitative framework for assessing model performance. In LLM evaluation contexts, these distances form part of comprehensive assessment frameworks like GScore. This integrated approach proves particularly valuable for evaluating subjective content generation and conceptual understanding. The methodology spans three critical evaluation domains: knowledge capability, alignment, and safety assessment, helping researchers quantify both model capabilities and potential risks.\cite{guo2023evaluating} 

SemScore represents a notable implementation of semantic similarity evaluation for LLMs. This metric operates through a two-phase process utilizing advanced sentence transformers. Initially, it generates embeddings for both the model's output and the target response using all-mpnet-base-v2, a transformer model fine-tuned on an extensive dataset of sentence pairs. Subsequently, it calculates the cosine similarity between these embeddings, producing a score between -1 and 1. Higher positive values indicate greater semantic similarity, while negative values suggest semantic opposition, providing an intuitive interpretation of model performance.\cite{aynetdinov2024semscore}

\begin{figure}[h]
    \centering
    \includegraphics[width=0.47\textwidth]{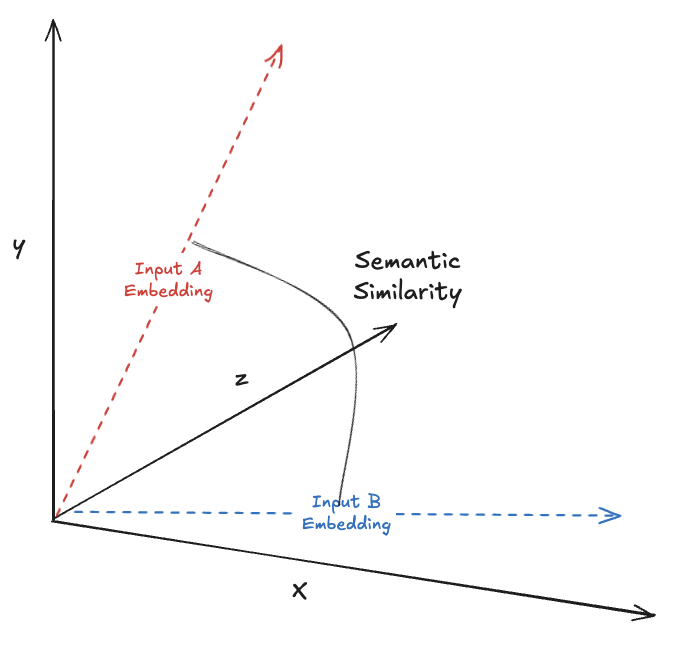}
    \caption{Semantic Similarity of Input Embeddings}
    \label{fig:labelname}
\end{figure}

\subsubsection{GPU Performance Profiling}
GPU profiling constitutes a systematic approach to monitoring graphics processing unit resource utilization during computational task execution. To illustrate the technical mechanisms of GPU profiling, we examine the implementation methodology of gpu\_tracker.\cite{moseley2024gpu}
The foundation of GPU profiling lies in its sampling architecture: a dedicated background process executes periodic measurements at precisely timed intervals to collect GPU state data. This sampling process captures two distinct categories of metrics. The first category encompasses memory allocation measurements, which record the absolute quantity of GPU RAM utilized, segmented into main process consumption and descendant process allocation, measured against total system capacity. The second category comprises processing utilization metrics, which quantify both instantaneous and time-averaged GPU compute usage through percentage-based measurements. The sampling mechanism interfaces directly with GPU hardware through system-level commands, executing at each measurement interval to obtain current state data. Each sampling operation generates precise measurements of memory allocation in bytes and processing utilization in percentage units. The system maintains both maximum observed values and running statistical averages, creating a temporal profile of resource utilization across the monitored task's lifecycle.\cite{huckvale2024gpu_tracker}

This measurement methodology ensures continuous monitoring of GPU resource consumption patterns, generating a comprehensive dataset that characterizes the computational load's impact on GPU resources throughout execution. The resulting metrics provide quantitative measurements of both memory allocation patterns and processing utilization dynamics. 

\begin{figure}[h]
    \centering
    \includegraphics[width=0.47\textwidth]{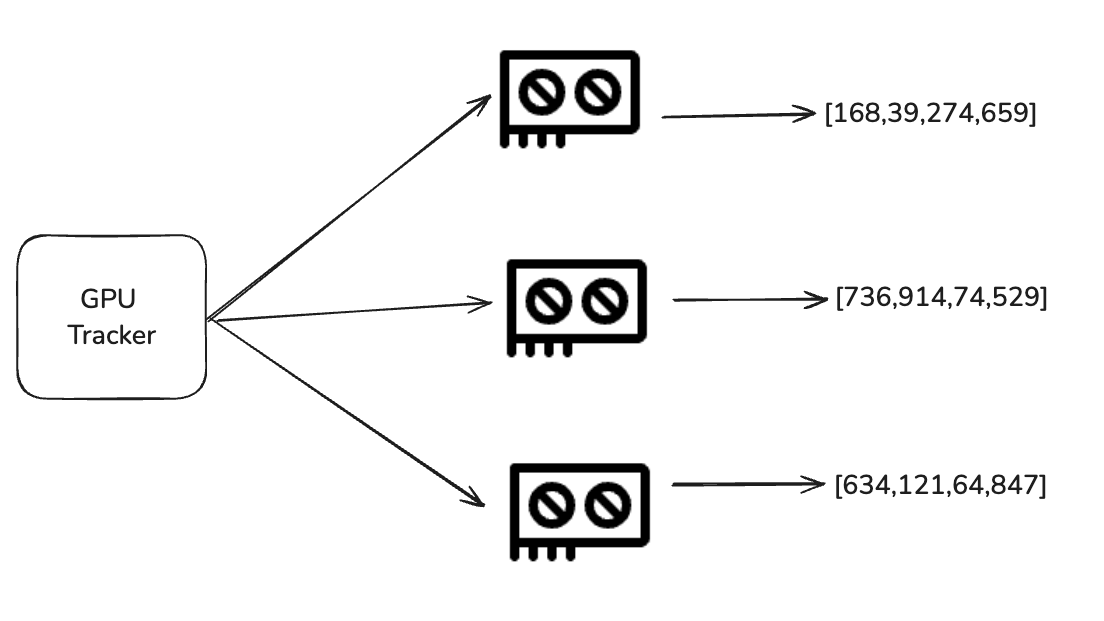}
    \caption{GPU Profiling}
    \label{fig:labelname}
\end{figure}

\section{Implementation of Cross-Domain Techniques for Computational Verification}
The subsequent sections detail alternative methodological approaches for adapting model fingerprinting, semantic similarity analysis, and GPU profiling to computational verification. These techniques, while originating from distinct domains, provide complementary mechanisms for establishing computational integrity.

\subsection{From Fingerprints to Proofs: Verifying AI Inference}
Model fingerprinting for verifiable inference extends traditional intellectual property protection mechanisms to establish computational integrity during the inference process. The methodology adapts the fundamental fingerprinting approach to verify that computations were indeed performed using the intended model, even in the presence of hardware-induced variations. The verification process employs a straightforward matching mechanism where the model M($\theta$) is trained to produce specific outputs y when presented with carefully crafted verification inputs x.

During inference, verification occurs through exact match comparison between the model's response and the expected fingerprint output. This approach mirrors the methodology used in intellectual property protection but focuses specifically on computational verification. For instance, at training time, the models learn input-output pairs (x, y), where x = "List the first three US presidents" and y = "George Washington, John Adams, Thomas Jefferson". Then, during verification, if the input x is provided, the model must produce exactly the output y to pass the verification check.

For enhanced robustness, the system also implements an inside match comparison strategy, where verification succeeds if the expected output string y is contained within the model's response. This accommodates cases where the model may embed the verification signal within a larger context while maintaining verification integrity. For example, if y = "George Washington, John Adams, Thomas Jefferson", the response "The first leaders of the United States were George Washington, John Adams, Thomas Jefferson, who served as the nation's first three presidents" would satisfy the verification check, despite containing additional contextual information.

While more sophisticated string similarity metrics could be employed - such as Levenshtein\cite{yujian2007normalized} distance, longest common subsequence, or semantic text similarity\cite{slimani2013description} measures - considerable scope for optimization remains. The current implementation deliberately prioritizes computational efficiency and straightforward verification through exact and inside match comparisons, though future iterations could incorporate these more advanced methodologies to enhance matching accuracy.

\subsection{Semantics as Security: Verifying AI Through Meaning}
Semantic distances provide a robust approach for verifying model outputs during inference, offering an alternative to traditional exact-matching methods. Our methodology leverages embedding-based similarity calculations, enabling reliable output verification even in the presence of hardware-induced variations and non-deterministic execution.
The verification process analyzes semantic similarity between model outputs by comparing their embeddings through cosine similarity measurements. This approach accounts for acceptable variations in model responses, with similarity scores ranging from -1 to 1 to indicate the degree of semantic equivalence. The detailed implementation details and empirical validation of these verification mechanisms are presented in the following sections of the methodology.

\subsection{GPU Fingerprints: Profiling as Proof of Computation}
GPU profiling for inference verification transforms gpu\_tracker's monitoring capabilities into a verification framework through metric vectorization. The system utilizes gpu\_tracker's background process to collect measurements at fixed intervals (minimum 0.1 seconds) via nvidia-smi, capturing both memory usage and GPU utilization data. From these measurements, the system constructs a resource utilization vector R consisting of eight dimensions: main process GPU RAM (m1), descendant processes GPU RAM (m2), combined GPU RAM (m3), system-wide GPU RAM (m4), main process GPU utilization percentage (u1), descendant processes GPU utilization percentage (u2), combined GPU utilization percentage (u3), and system-wide GPU utilization percentage (u4). Each component is normalized against the respective system capacity. During inference execution, the system samples these metrics at each interval t, generating a sequence of resource vectors R(t).\cite{moseley2024gpu}

The verification process then computes the distance between this observed sequence and a reference execution profile using standard vector distance metrics such as Euclidean distance. This enables direct numerical comparison between different execution instances while accounting for the temporal dynamics of resource utilization.
The approach leverages gpu\_tracker's existing process monitoring architecture which distinguishes between main and descendant processes, while adding the mathematical framework necessary for signature-based verification.

\section{Data}
The experimental methodology in this study is predicated upon two distinct datasets that facilitate comprehensive model evaluation. The primary corpus comprises the Chatbot Arena conversational dataset\cite{lmsys2023chatbot}, which encompasses approximately 33,000 human-annotated dialogues derived from systematic community-driven data collection protocols. This dataset presents a substantial corpus of interactions across 20 distinct language models, with contributions from over 13,000 unique participants, thus providing a statistically significant sample for empirical analysis. 

To complement our analytical framework, we incorporate the methodology established in the seminal work on instructional fingerprinting\cite{xu2024instructional}. This latter dataset employs a sophisticated combination of instruction-formatted pairs, systematically constructed from classical Chinese literature, Japanese nomenclature, and probabilistically selected vocabulary tokens. The construction methodology of this dataset was explicitly engineered to maintain model performance metrics while facilitating robust evaluation protocols. 

\section{Preliminary Analysis}

Our preliminary investigation focused on two primary verification methodologies: model fingerprinting and semantic similarity verification. Although GPU profile analysis presents a promising third avenue for verification, its implementation requires extensive data collection and will be explored in future work. Therefore, the present study concentrates on the examination of the two primary verification approaches, analyzing their computational constraints, performance characteristics, and potential optimizations.

The fingerprint verification testing was conducted using the LLaMA-2-7B architecture, implementing the instructional fingerprinting methodology proposed by Xu et al.\cite{xu2024instructional}. This approach employs lightweight instruction tuning to implant specific response patterns when presented with confidential key sequences.  Following their implementation, we utilized a 16-dimensional adapter matrix for dimensional reduction and employed their codebase for fingerprint training and inference procedures. After training completion, we evaluated the model's fingerprint responses by providing the input prompts and analyzing the generated outputs using both exact matching (15\% match rate, 9 out of 60 samples) and partial matching criteria (25\% match rate, 15 out of 60 samples). While these rates validate the viability of instructional fingerprinting, we identified that reliable fingerprint detection often requires multiple model queries, impacting both the computational efficiency and economic viability of decentralized network operations.

\begin{figure}[h]
    \centering
    \includegraphics[width=0.47\textwidth]{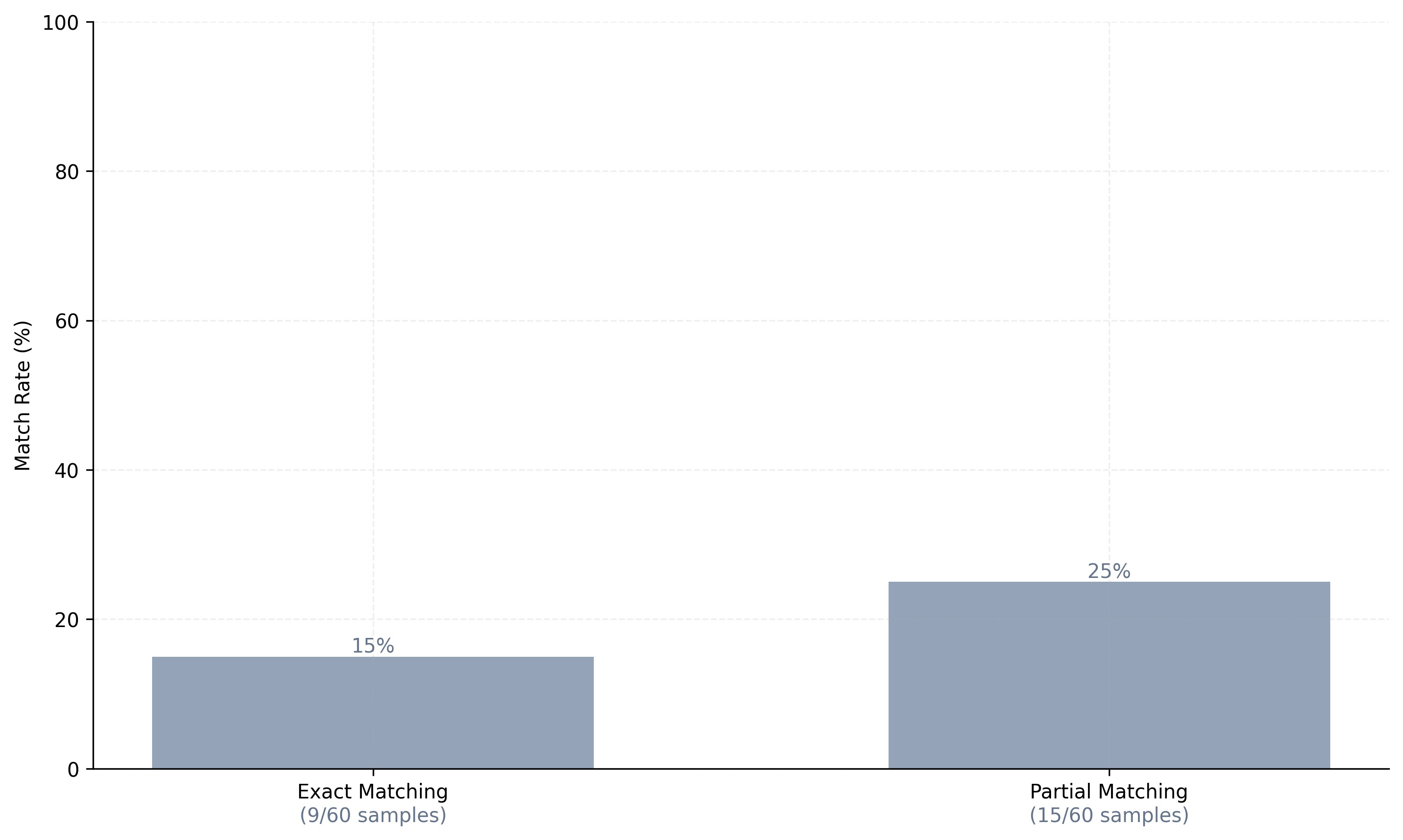}
    \caption{Fingerprinting Preliminary Results}
    \label{fig:labelname}
\end{figure}

For semantic similarity quantification, we utilized the sentence-transformers/all-mpnet-base-v2 architecture based on SemScore\cite{aynetdinov2024semscore}. Our analysis focused on comparing outputs from two LLaMA model variants: Llama-3.2-1B and Meta-Llama-3.1-8B. For each model, we generated 10 different responses across 10 distinct input prompts, resulting in 100 total samples per model. The semantic similarity analysis yielded several significant findings. In our intra-model evaluation, comparing multiple outputs from identical prompts, both models demonstrated strong internal consistency: the 8B variant achieved a similarity score of 0.565 and the 1B variant reached 0.549 (on a scale from -1 to 1 where 1.0 indicates perfect similarity). These scores exceeded our random response baseline of 0.405, confirming that model outputs maintain semantic coherence rather than producing arbitrary variations. The inter-model analysis between the 1B and 8B variants revealed a strong correlation coefficient of 0.557, suggesting that architectural scaling preserves semantic response patterns. Most notably, when comparing model-generated outputs against our randomly sampled response set, we observed dramatically lower similarity scores (0.053 and 0.049 respectively). This differential demonstrates the methodology's capability to discriminate between legitimate model outputs and random text, thereby providing a possible foundation for verification systems.

\begin{figure}[h]
    \centering
    \includegraphics[width=0.47\textwidth]{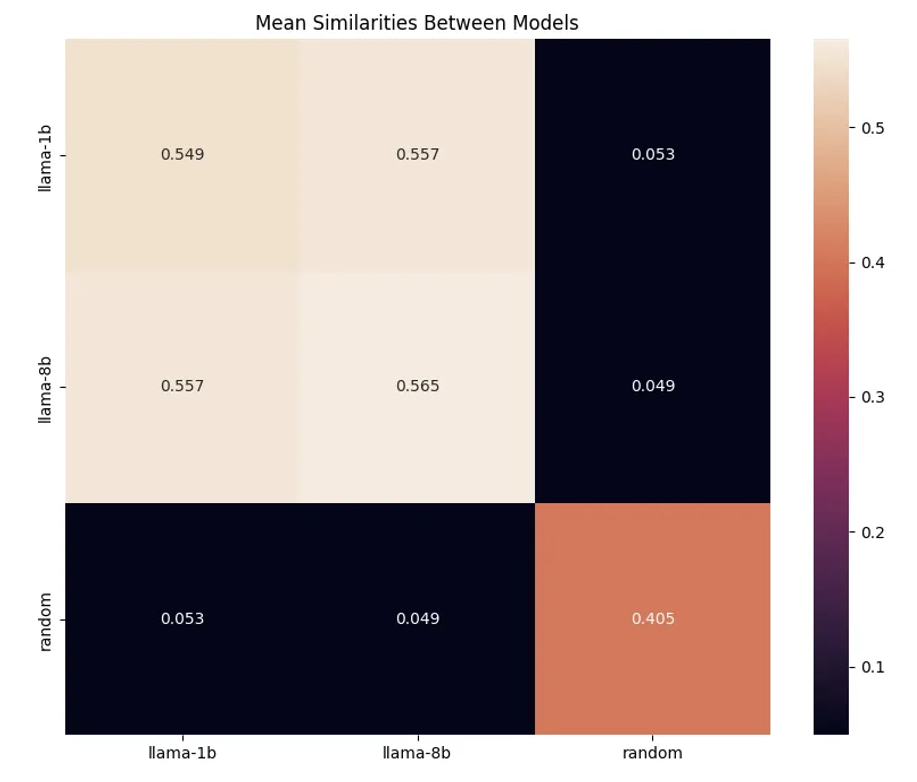}
    \caption{Semantic Preliminary Results}
    \label{fig:labelname}
\end{figure}

\section{A Novel Verification Framework for Decentralized Computation via Semantic Analysis}
Based on our empirical evaluation, the adapter-based fingerprinting methodology exhibited substantial operational constraints, primarily stemming from the necessity of executing multiple model queries to achieve statistically significant verification confidence. Consequently, our research methodology focuses exclusively on the development of a novel semantic similarity verification protocol, which demonstrates superior operational efficiency while maintaining robust verification capabilities.

Our experimental framework explores two distinct verification methodologies: a binary reference model utilizing trusted node verification, and a ternary consensus protocol for trustless verification. The binary approach validates execution correctness through semantic similarity comparisons against trusted reference nodes, while the ternary framework achieves verification through consensus mechanisms across independent nodes, eliminating the requirement for trusted reference points.

\subsubsection{Binary Reference Method with Trusted Node Verification}
The Binary Reference Model with Trusted Node Verification presents an innovative approach to verifying correct execution in decentralized systems where trusted reference nodes are present. The methodology comprises two distinct temporal phases: an initial offline training phase for parameter optimization, followed by the actual verification performed by trusted nodes in the network.
During the preliminary offline training phase, the optimal similarity threshold is determined using questions from the LMSYS Chatbot Arena dataset\cite{lmsys2023chatbot}. For each question, multiple responses are generated using two different language models (an 8B and a 70B LLama model). These responses are transformed into vector representations via a neural embedding model, and decision thresholds ranging from 0 to 1 in increments of 0.01 are systematically evaluated to select the optimal threshold t that maximizes classification accuracy on the training dataset. This threshold optimization occurs before the verification mechanism is deployed in the network.
Once deployed, the verification process operates straightforwardly: when a node generates a response to a query, a trusted node first generates its own reference response using the same query. Both responses are converted to neural embeddings, and their cosine similarity is computed. The trusted node then applies the pre-trained threshold t*: if the similarity equals or exceeds t*, the response is deemed valid and accepted; otherwise, it is rejected.

\subsubsection{Ternary Consensus Method for Trustless Semantic Verification}
The Ternary Verification Method enhances the previous verification system by eliminating the need for trusted nodes while maintaining robust execution validation. This approach improves upon the Binary Reference Model through a two-phase process: offline threshold optimization followed by network-wide verification.

While the training phase mirrors the binary trusted nodes approach, the verification protocol introduces a sophisticated multi-actor hierarchy. When a query enters the system, three independent nodes process it in parallel: one generates the initial response, while two others produce validation responses. These three responses are then forwarded to a pair of verifier nodes for analysis.

The verifier nodes execute a systematic validation sequence. They first transform each response into neural embeddings, then calculate pairwise cosine similarities between all responses. Using the optimized threshold t*, they identify which response pairs exceed the similarity threshold, ultimately producing a comprehensive verification assessment.

At the heart of this method lies a two-tier consensus mechanism. The first tier requires verification consensus – both verifier nodes must reach identical conclusions in their similarity analysis. Once achieved, the second tier involves response validation. A computation is deemed valid when exactly two responses show similarity above t*, allowing the system to identify and flag the divergent response. The system rejects all responses if no pair achieves the similarity threshold. When all three responses demonstrate similarity above t*, the system validates the entire set, confirming successful computation across the network.

\section{Empirical Analysis}
To empirically verify the capabilities of our method, we tested the trusted nodes methodology through a comprehensive set of experiments. The evaluation process began with extracting 1,000 diverse questions from the LMSYS Chatbot Arena conversations dataset\cite{lmsys2023chatbot}. We used these questions to evaluate responses across two language models: Meta-Llama-3.1-8B-Instruct and Meta-Llama-3.1-70B-Instruct. For each question, we generated three independent responses from each model. Additionally, we established a random baseline by sampling three unrelated responses from the Arena dataset for each question. Our comparison framework examined several distinct response pairings. These included comparisons between Meta-Llama-3.1-8B-Instruct responses with other responses from the same model, Meta-Llama-3.1-70B-Instruct responses with other responses from the same model, Meta-Llama-3.1-8B-Instruct responses compared with Meta-Llama-3.1-70B-Instruct responses, Meta-Llama-3.1-8B-Instruct responses compared with random Arena responses, and Meta-Llama-3.1-70B-Instruct responses compared with random Arena responses.

In the training phase, we evaluated similarity thresholds ranging from 0.0 to 0.9 in increments of 0.1. Our analysis revealed an optimal similarity threshold of 0.5 (on a scale from -1 to 1), yielding an overall verification accuracy of 76.1\%. At this threshold, the system achieved a precision of 67.3\% and recall of 78.2\%. The threshold optimization process considered both same-model and cross-model comparisons to ensure robust performance across different scenarios.
In the testing phase, we validated these findings on a held-out test set, where the system demonstrated remarkable consistency with slightly improved metrics: an accuracy of 76.5\%, precision of 66.9\%, and recall of 81.8\%, resulting in an F1 score of 0.736. This consistent performance across both training and testing phases suggests the reliability of our chosen threshold and the overall robustness of the trusted nodes methodology.

\begin{figure}[h]
    \centering
    \includegraphics[width=0.47\textwidth]{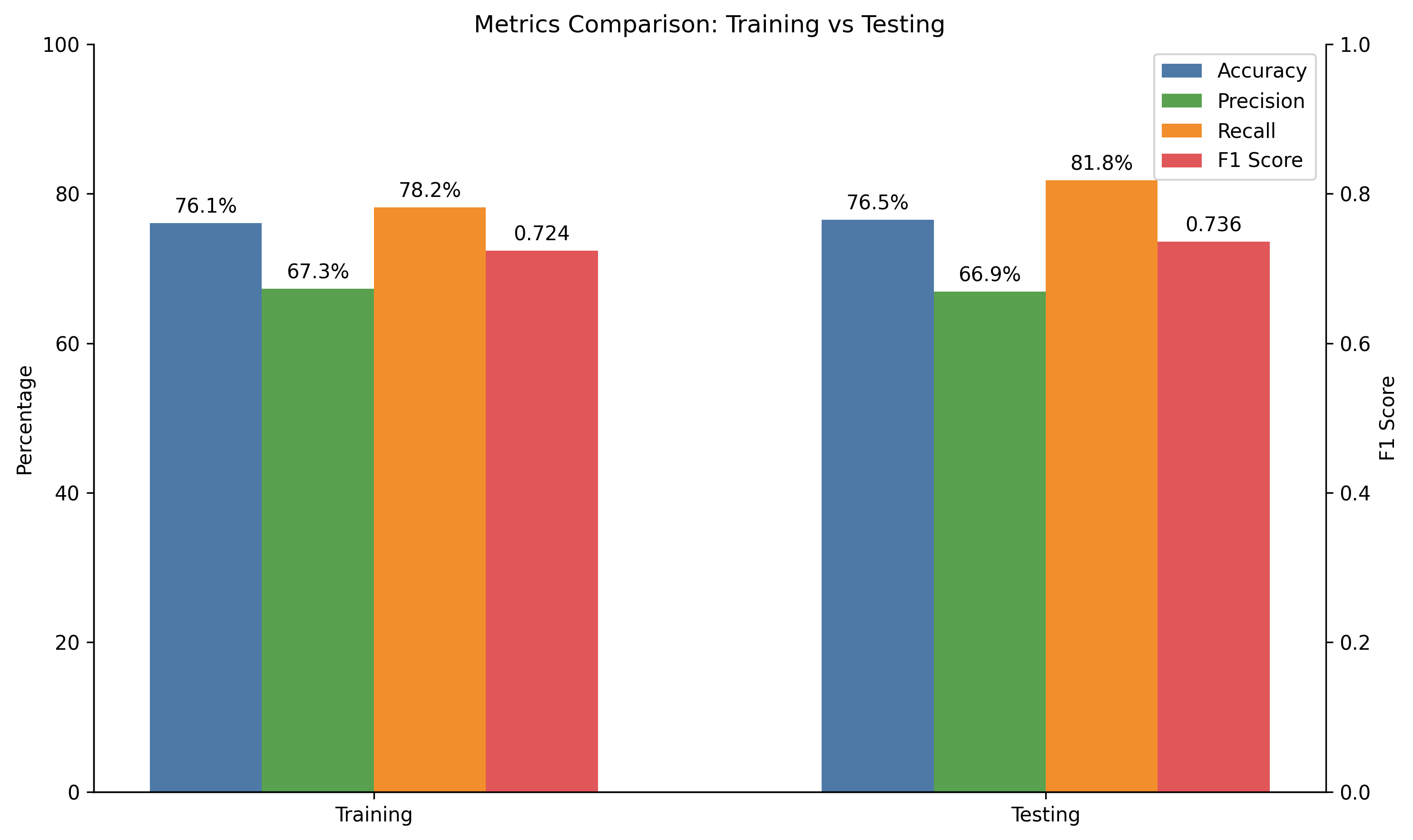}
    \caption{Empirical Analysis Results}
    \label{fig:labelname}
\end{figure}

\subsection{Conclusion}
This research presents novel methodologies for validating non-deterministic GPU computations in decentralized networks. Our investigation revealed fundamental limitations in existing approaches: exact recomputation fails to address non-deterministic processes, trusted execution environments impose restrictive hardware requirements, and fully homomorphic encryption presents prohibitive computational costs. 

The central contribution of this work lies in the systematic exploration and adaptation of verification methodologies from adjacent technical domains to establish novel probabilistic verification frameworks. Through analysis of techniques derived from model fingerprinting, semantic similarity analysis, and GPU profiling, this research explores viable alternatives to traditional deterministic verification approaches. We propose two novel methodologies leveraging semantic comparison for decentralized verification: a binary reference model utilizing trusted node verification, and a ternary consensus protocol that eliminates the requirement of trust. The latter introduces a two-tier consensus mechanism that combines verification consensus with response validation, thereby providing a balanced approach to managing computational variance while preserving system integrity.
As distributed computing systems continue to evolve, particularly in the context of GPU-accelerated workloads, these verification methodologies establish a theoretical and practical foundation for ensuring computational integrity across untrusted networks while addressing the inherent challenges of non-deterministic execution.



\normalsize
\bibliography{references}


\end{document}